\documentclass[10pt,conference]{IEEEtran}
\IEEEoverridecommandlockouts
\usepackage{cite}
\usepackage{amsmath,amssymb,amsfonts}
\usepackage{algorithmic}
\usepackage{graphicx}
\usepackage{textcomp}
\usepackage{xcolor}

\usepackage{graphicx}
\usepackage{subfigure}
\usepackage{multirow}
\usepackage{multicol}
\usepackage[breakable]{tcolorbox}
\usepackage{pifont}
\usepackage{enumitem}
\usepackage{amsmath}
\usepackage{hyperref}
\usepackage{bbm}

\usepackage{xspace}
\newcommand{\ie}{{\emph{i.e.}}\xspace}
\newcommand{\etc}{etc.}
\newcommand{\eg}{{\emph{e.g.}}\xspace}

\usepackage[numbers,sort&compress]{natbib}
\usepackage[normalem]{ulem}
\useunder{\uline}{\ul}{}
\usepackage{threeparttable}
\usepackage{makecell}
\usepackage{tcolorbox}
\usepackage{booktabs}

\usepackage{listings}

\usepackage{tikz}
\usepackage{textcomp}

\def\BibTeX{{\rm B\kern-.05em{\sc i\kern-.025em b}\kern-.08em
    T\kern-.1667em\lower.7ex\hbox{E}\kern-.125emX}}

\newcommand*{\circled}[1]{\lower.7ex\hbox{\tikz\draw (0pt, 0pt)    circle (.3em) node {\makebox[1em][c]{\tiny #1}};}}
\newcommand{\paragraphwithoutdot}[1]{\vskip 0.01in \noindent {\bf #1}}
\renewcommand{\paragraph}[1]{\vskip 0.01in \noindent {\bf #1.}}
\newcommand{\baseline}[1]{\vskip 0.01in \noindent {\textit{#1.}}}
\newcommand{\baselinewithoutdot}[1]{\vskip 0.01in \noindent {\textit{#1}}}
\newcommand{\baselineinline}[1]{{\textit{#1}}}

\newcommand{\definenode}[1]{{\fontfamily{\ttdefault}\selectfont #1}}

\newcommand{\citea}[1]{\citeauthor{#1}~\cite{#1}}

\newcommand{\greysum}[1]{}

\begin{document}

\title{Implant Global and Local Hierarchy Information to Sequence based Code Representation Models
}
\makeatletter 
\newcommand{\linebreakand}{%
  \end{@IEEEauthorhalign}
  \hfill\mbox{}\par
  \mbox{}\hfill\begin{@IEEEauthorhalign}
}
\makeatother 

\author{\IEEEauthorblockN{Kechi Zhang\textsuperscript{†}}
\IEEEauthorblockA{
Key Lab of High Confidence Software \\
Technology, MoE (Peking University) \\
Beijing, China \\
zhangkechi@pku.edu.cn}
\and
\IEEEauthorblockN{Zhuo Li\textsuperscript{†} \thanks{† The two authors share equal contribution.}}
\IEEEauthorblockA{
Key Lab of High Confidence Software \\
Technology, MoE (Peking University) \\
Beijing, China \\
lizhmq@pku.edu.cn}

\linebreakand

\IEEEauthorblockN{Zhi Jin*}
\IEEEauthorblockA{
Key Lab of High Confidence Software \\
Technology, MoE (Peking University) \\
Beijing, China \\
zhijin@pku.edu.cn}
\and
\IEEEauthorblockN{Ge Li* \thanks{* Corresponding authors}}
\IEEEauthorblockA{
Key Lab of High Confidence Software \\
Technology, MoE (Peking University) \\
Beijing, China \\
lige@pku.edu.cn}
}

\maketitle
\begin{abstract}
Source code representation with deep learning techniques is an important research field. 
There have been many studies that learn sequential or structural information for code representation.
But sequence-based models and non-sequence-models both have their limitations.
Researchers attempt to incorporate structural information to sequence-based models, but they only mine part of token-level hierarchical structure information.
In this paper, we analyze how the complete hierarchical structure influences the tokens in code sequences and abstract this influence as a property of code tokens called hierarchical embedding. The hierarchical embedding is further divided into statement-level global hierarchy and token-level local hierarchy. 
Furthermore, we propose the Hierarchy Transformer (HiT), a simple but effective sequence model to incorporate the complete hierarchical embeddings of source code into a Transformer model.
We demonstrate the effectiveness of hierarchical embedding on learning code structure with an experiment on variable scope detection task. Further evaluation shows that HiT outperforms SOTA baseline models and show stable training efficiency on three source code-related tasks involving classification and generation tasks across 8 different datasets.

\end{abstract}

\section{Introduction}
\label{sec:intro}










%


Code representation is a hot research topic in software engineering (SE) and machine learning (ML) fields. Machine learning for code representation learning aims to convert programs of different formats (sequential formats such as token sequences, structural formats such as abstract syntax trees, dependency graphs, \etc) into vectorized semantic embeddings. These representation vectors can be applied on many downstream tasks,
such as code classification \cite{tbcnn}, type inference \cite{allamanis2020typilus}, code summarization \cite{code2vec,alon2018code2seq,IyerKCZ16codenn,hu2018deep}, \etc

\begin{figure}[t]
 \subfigure[Hierarchical location affects the operational semantics of a statement] {
  \label{fig:statement}     
  \includegraphics[width=\columnwidth]{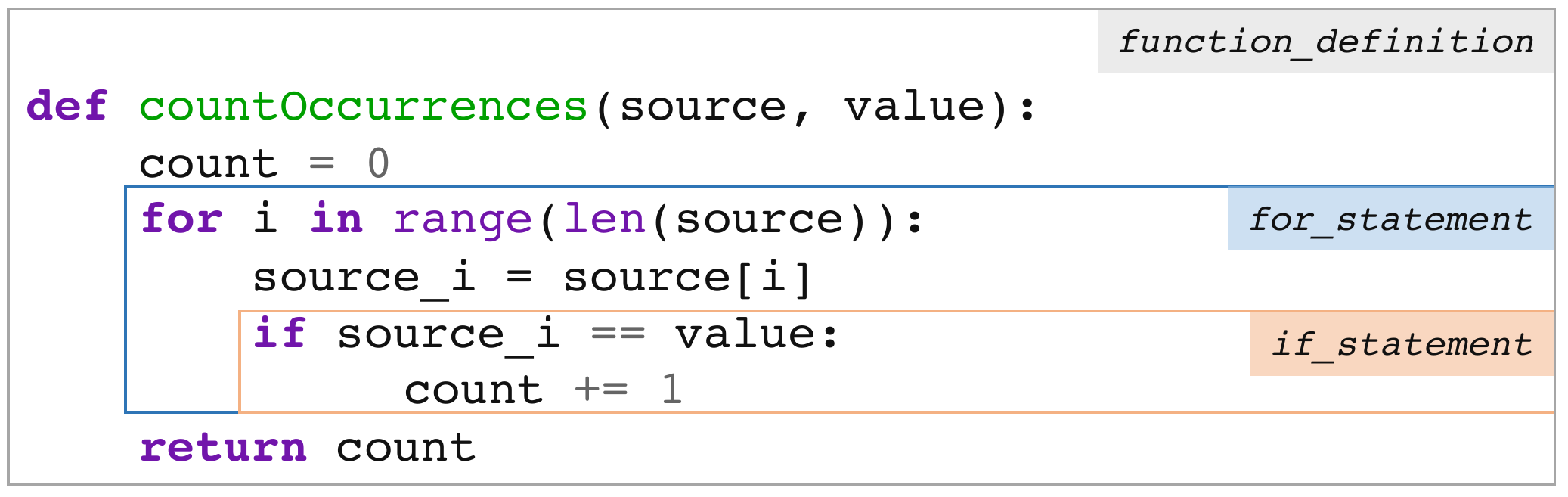}  
  }

 \subfigure[Hierarchy information within a statement affects the semantics of a token.] {
  \label{fig:locality}     
  \includegraphics[width=\columnwidth]{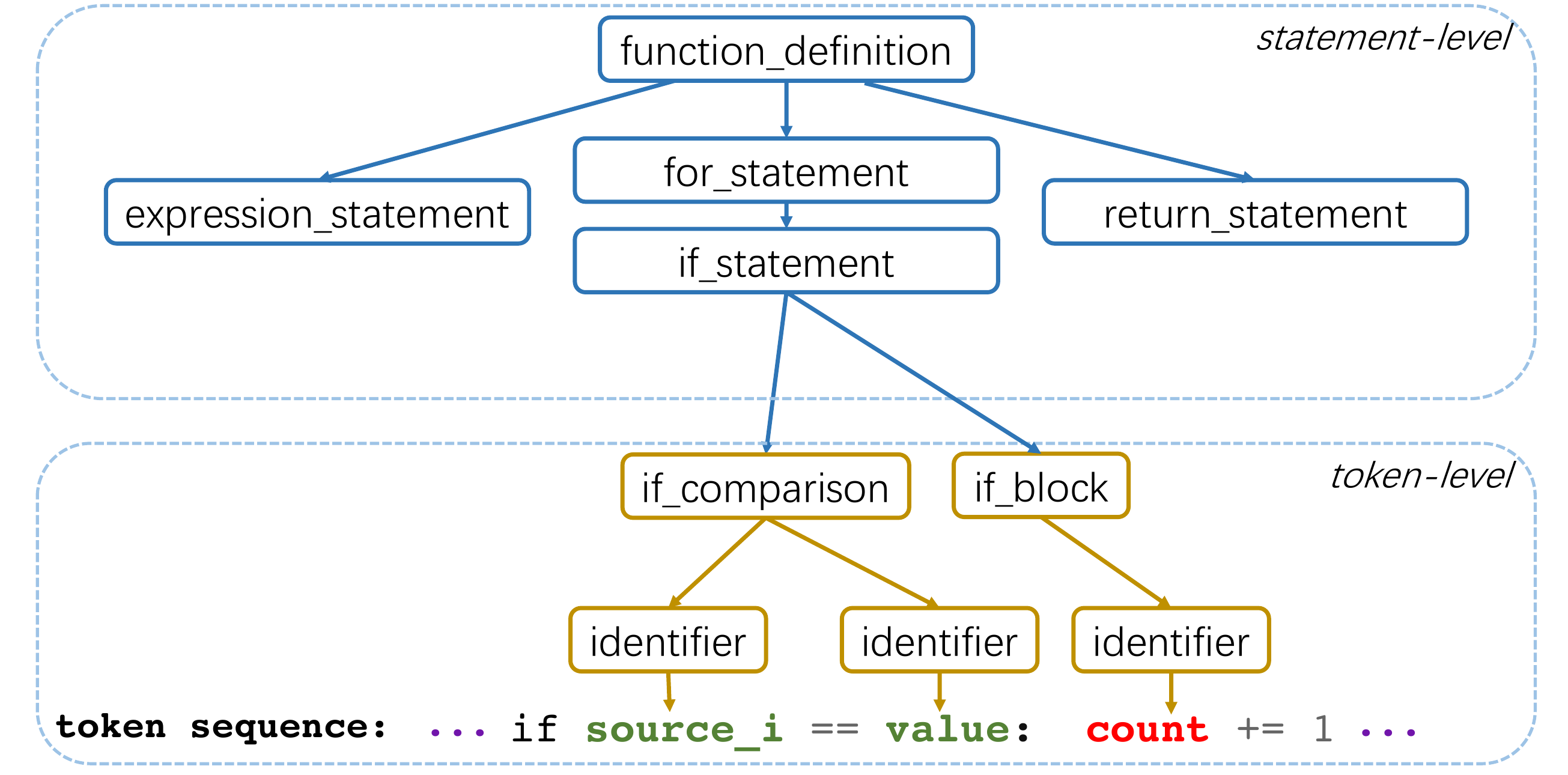}  
  }
\caption{An illustrative example of the hierarchy information in source code. The hierarchy of a token refers to the hierarchical location of its statement (\textit{global hierarchy}) and the local component of the token in the statement (\textit{local hierarchy}). The token \definenode{source\_i} and \definenode{value} are both \definenode{identifier} in \definenode{if\_comparison} marked in green in (b) and they are more closely related.}
\label{fig:intro_locality}
\vspace{-15pt}
\end{figure}

Most existing code representation methods can be divided into two categories: Sequence-based models \cite{AllamanisBDS18naturalness, cbert} are skilled at processing sequence order information with long-term dependency.
But they are sub-optimal for capturing structural information \cite{tbcnn,icseZhangWZ0WL19astnn}. Non-sequence-based models, such as tree-based (ASTNN \cite{icseZhangWZ0WL19astnn}, TBCNN \cite{tbcnn}) or graph-based models (GGNN \cite{li2015ggnn}), focus on encoding structural information, but sacrifice the advantages of sequence models or suffer from narrow receptive fields. These methods predominantly leverage either sequential or structural information of source code, and ignore the combination of the two modal information.

Recently, some studies jointly learn both sequential and structural information for code representation in sequence-based models. \citea{hellendoorn2019global} proposed \textit{GREAT}, a Transformer model using the relative positional encoding to bias the attention using edges of the data flow graph and control flow graph. \citea{ZugnerKCLG21codetrans} proposed \textit{CodeTransformer}, which combines distances computed on the AST in the self-attention operation. However, these methods use limited information of code structure, such as node distances on the program tree or graph.
They focus on modeling structure as a relation between tokens with attention mechanism and overlook the full impact of hierarchical structure information, which weakens the structural information of source code.

In this work, we take a step to explore how the complete hierarchical structure influence the tokens in source code sequence and abstract this influence as a property of code tokens called hierarchical embedding.
We further divide this influence into two aspects:
\ding{182} hierarchical embedding of statements (named as \textit{\bf global hierarchy}).  The hierarchical embedding will affect the operational semantics of a statement. For example, in Figure \ref{fig:intro_locality}, the statement \definenode{count += 1} is written in the \definenode{for-block} and possibly to be executed more than once, while the statement \definenode{count = 0} will be executed only once in each function call to \definenode{countOccurrences}.
\ding{183} hierarchical embedding of tokens within a statement (named as \textit{\bf local hierarchy}). Tokens are in different components in a statement, which will affect the semantics of each token. \eg, as shown in Figure \ref{fig:locality},
the three identifiers, \textit{source\_i},  \textit{value}, and \textit{count}, in the \definenode{if-statement}, are references to variables in the function. However, the first two identifiers within \definenode{if-comparison} component (marked in green color in the figure) have more similar semantics than \textit{count} because they appear in the same comparison expression.



To validate our intuition of modeling the global and local hierarchies with code sequences, we propose to unify hierarchical structure into the code sequence in a concise Transformer format. We name the network \textbf{Hi}erarchy \textbf{T}ransformer (HiT), a simple but effective model that can achieve a preferable balance between efficiency and effectiveness. The HiT model consists of two key components: a Transformer-based hierarchy encoder that learns the representation of the hierarchy information, and a Transformer-based sequence encoder that fuses the hierarchy information and token sequence information.
Specifically, the hierarchy information is represented by the root-to-leaf paths in the code syntax tree and encoded by the Transformer-based hierarchy encoder.


We conduct an empirical study to investigate the impact of the global and local hierarchy.
The experiment proves the effectiveness of the two different types of hierarchical embeddings. It shows that with the cost of a small number of additional parameters, our approach significantly enhances the performance of sequence-based code representation models and achieves stable training efficiency.
We also design a variable scope detection task to show our model can well learn the scope information in global hierarchy and represent the relationship between sequence and structure information of source code.
We further evaluate our approach on three tasks: code classification, clone detection, and method name prediction, with 8 different datasets from different domains. These tasks include classification and generation.
The results show that our approach outperforms existing code representation models and other state-of-the-art models. 
It indicates the benefits of our approach for aligning the complete hierarchical embedding with code tokens for source code understanding.

The contributions of this paper are summarized as follows.

\begin{itemize}[leftmargin=*]
    \item We analyze how the complete hierarchical structure influences tokens in code sequences and abstract this influence as a property of code tokens called hierarchical embedding.
    \item We propose HiT, a simple but effective model to incorporate the hierarchical embedding into Transformer. Through experimental analysis, we demonstrate our approach can well represent the scope information in global hierarchy. Our empirical study shows that global and local hierarchical information are essential for code representation models, while existing jointly learning models ignore the former.
    \item We evaluate our approach on 3 source code-related tasks with 8 different datasets, involving classification tasks and generation tasks on source code. Our experimental results prove that aligning the complete hierarchical embedding with code tokens is effective for learning representations for programs with stable training efficiency. \footnote{Code and data are open-sourced on Github: \url{https://github.com/zkcpku/HiT-hierarchy-transformer}}
\end{itemize}

\section{Related Work}
\label{sec:related}
\subsection{Sequence-based Code Representation }

Code representation learning is a hot research topic in software engineering and machine learning fields. 
Among various representation approaches, sequence models are the most mainstream code representation models, based on the concept of ``naturalness`` 
\cite{naturalness2012, AllamanisBDS18naturalness, cbert, yangmengfei2022intelligent}, 
which argues that programming languages are usually simple, repetitive, and can be understood through the same approaches used in natural language processing. 
Sequence models are efficient and effective in processing the code token sequence, and have been applied across many SE tasks \cite{IyerKCZ16codenn,cai2020tag,AhmadCRC20trans,allamanis2016convolutional,Liu0BKKKKT19spot,NguyenPLN20MNire,Wang0LM21cognac,liu22GTNM}.
Recently, sequence-based pre-trained models \cite{FengGTDFGS0LJZ20Codebert, LuGRHSBCDJTLZSZ21codexglue,10.1145/3540250.3549081}, such as CodeBERT, have achieved success in SE tasks, demonstrating the power of sequence-based models. Due to their large-scale parameters and massive pre-training data, We did not consider pre-trained models as baseline models in this work.

There are also researchers trying to encode structural information with sequence models \cite{code2vec,alon2018code2seq,hu2018deep,SPTCode,li2023skcoder}. 
Some leverage the program AST to model the structure in source code and represent source code as a set of leaf-to-leaf paths over ASTs \cite{alon2018code2seq}.
Others use the flattened (AST) node sequence as input to model the structure in the source code.
However, leaf-to-leaf paths would scrap the code sequence information.
Using the flattened node sequences to encode tree structures makes the entire sequence representation significantly longer, and code tokens are interspersed with other non-terminal tree nodes.
These approaches weaken the “naturalness” of the source code context. We declare that combining the “naturalness” and hierarchical structure information in the code sequence is essential. In this paper, we follow the research line of incorporating hierarchical structure into the code sequence representation model.

\subsection{Non-Sequence-based Code Representation}

Programs contain extensive structure information. Therefore, recent studies explore using tree-based \cite{tbcnn,icseZhangWZ0WL19astnn, aaaiBuiYJ21treecaps, DBLP:conf/icml/0002SLY20,Wang2022LearningPR} and graph-based \cite{allamanis2018learning, fernandes2018structured,yin2018learning,wang2020detecting, DBLP:journals/pacmpl/WangWGW20,10.1145/3524610.3527905} models for code representation models.
Although tree-based and graph-based models can directly capture structural information of source code, they are generally less efficient than sequence models. They require complex data preprocessing designed for particular languages. In the input tree or graph, the number of nodes in the receptive field of each node grows exponentially, which leads to models not being able to understand the complete sequence information well \cite{iclr0002Y21bottleneck}. In our experiments, we include graph-based and tree-based models as our baselines.

Recently, some studies have also integrated structure information of source code into sequence-based models \cite{hellendoorn2019global,DBLP:conf/icse/LeClairJM19, ZugnerKCLG21codetrans, DBLP:conf/icse/KimZT021, graphcodebert}. Some of these studies are designed for code representation tasks: 
\citea{hellendoorn2019global} proposed Graph Relation Embedding Attention Transformer (GREAT), which biases Transformer with relational information from graph edge types, and achieves good performances on variable misuse task. \citea{ZugnerKCLG21codetrans} proposed CodeTransformer, which computes pairwise distances on AST and integrates multiple relations into the attention module.
However, these methods jointly learn source code's sequential and structural information with a simplified token-level hierarchical structure, such as node distances on the program tree or graph. They overlook the full impact of hierarchical structure information on the code sequence, and we will give a deep analysis of the hierarchy information contained in the code structure.



\section{Analysis of Hierarchy Information}
\label{sec:motivation}
In this section, we analyze how the complete hierarchy information influences the tokens in the source code sequence and motivating examples. 
Generally, for programs, we tokenize the source code to get the token sequences and encode them with a sequence model with hierarchical embeddings.
Hierarchical embedding can be regarded as a semantic property of each token in the sequence.
We further divide the semantics in the hierarchical embedding into two levels: hierarchical embedding of tokens within a statement (named as \textit{\bf local hierarchy}) and hierarchical embedding of statements (named as \textit{\bf global hierarchy}).
We analyze the token-level and statement-level semantics introduced by each part of the hierarchical embedding. 

\subsection{Understanding Token-level Semantics with Local Hierarchy}

The semantics of the token is related to the local hierarchy.
Simply encoding a token with only the token embedding will lose local structural information.
Specifically, the same tokens may have different semantics, but given the same token embedding. On the contrary, different tokens that appear in similar contexts may have similar semantics.

For example, in Figure \ref{fig:reused_ij}, the variable \definenode{i} in the first program is used as a loop variable. In contrast, the variable \definenode{i} in the quick sort program represents the partition index. Although both programs define and use the variable \definenode{i}, they express different meanings with it.
Furthermore, in the first program in Figure \ref{fig:reused_ij}, there is another variable \definenode{j} used as a loop variable.
We can discover that the variable \definenode{i} and \definenode{j} in the first program have similar semantics, which is different from the meaning of variable \definenode{i} in the second program.
However, the embedding layer gives the same representation to \definenode{i} and a different representation to \definenode{j}.
Another example is shown in Figure \ref{fig:reused}. In these two lines of python code, the token appears three times, indicating a module import, a module reference, and a module attribute, respectively. It is hard to exploit this implicit semantic difference for a traditional sequence model.
Thus we believe that it is essential to model the structure of the local hierarchy to understand the meaning of tokens.

We revisit existing joint learning models for code representation \cite{hellendoorn2019global,ZugnerKCLG21codetrans} and find that most of them focus on encoding relations between tokens in code sequences, such as node distances on the program tree/graph. They are skilled at encoding the token-level semantics and yield good performance. But our further analysis shows that hierarchical structural information contains more than token-level information.


\begin{figure}[t]
  
  \includegraphics[width=0.95\columnwidth]{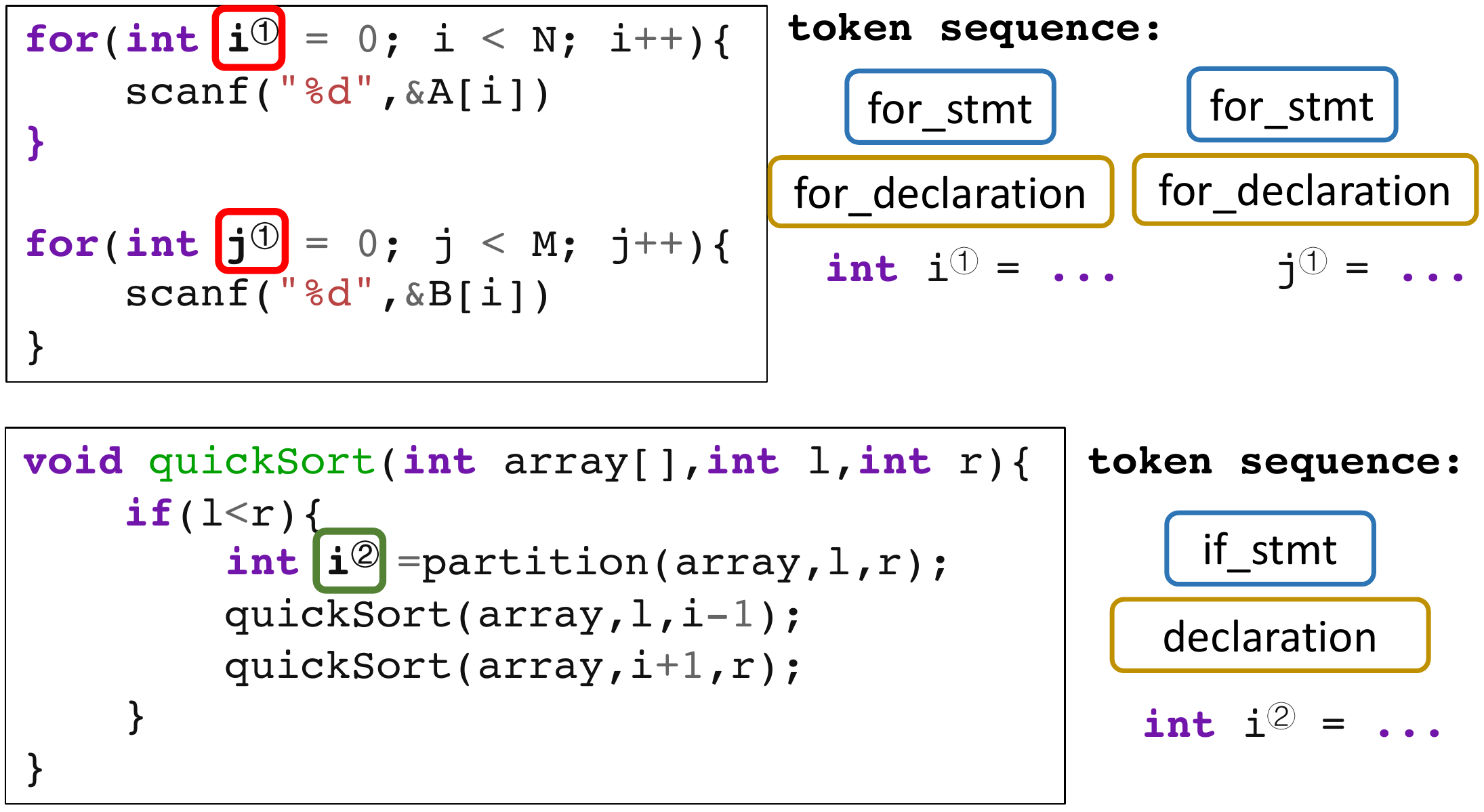}  
\caption{An illustrative example of thse tokens in different contexts. The same token may have different semantics ( $i^{\tiny \textcircled{1}}$ and $i^{\tiny \textcircled{2}}$ in two programs), and different tokens with similar context may have similar semantics ( $i^{\tiny \textcircled{1}}$ and $j^{\tiny \textcircled{1}}$ in the first program) }


\label{fig:reused_ij}
\vspace{-10pt}
\end{figure}

\begin{figure}[t]
  
\centering
  \includegraphics[width=0.55\columnwidth]{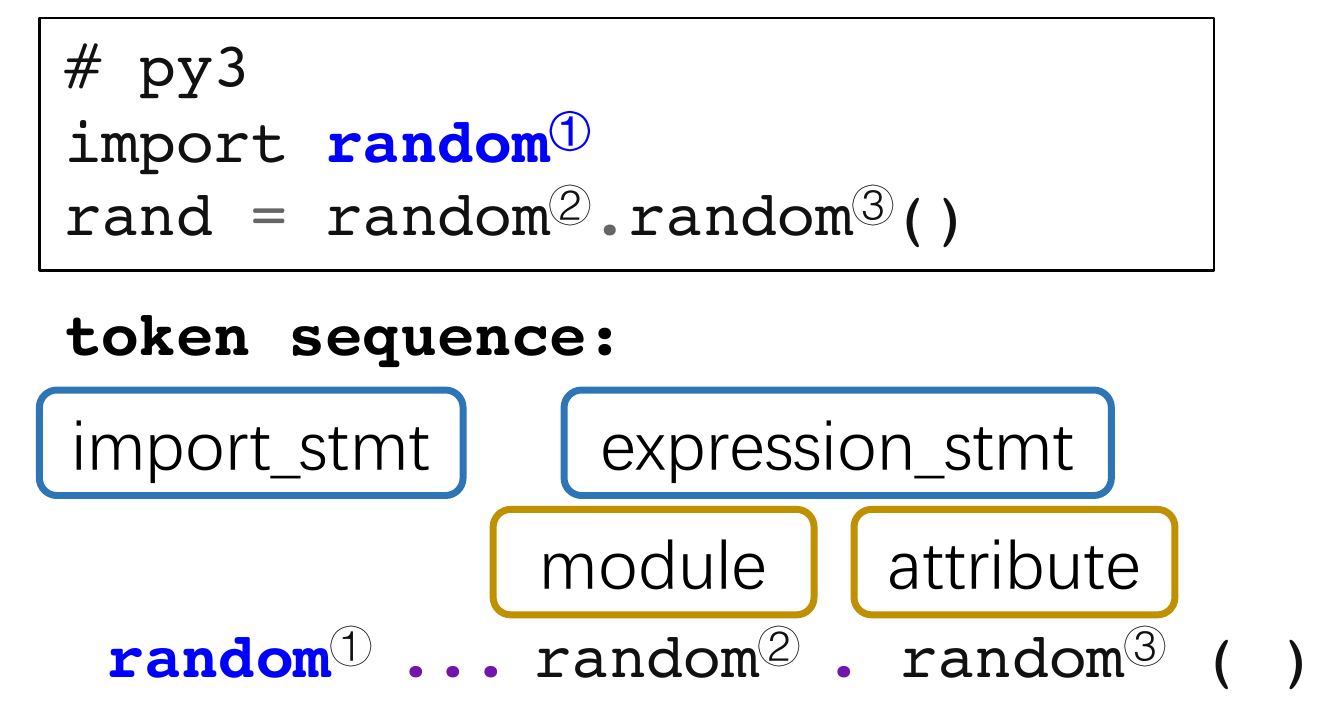}  
\caption{An illustrative example of the repeated tokens in different statements. The repeated token \definenode{random} represents different semantics.}
\label{fig:reused}
\vspace{-20pt}
\end{figure}

\subsection{Understanding Statement-level Semantics with Global Hierarchy}

To illustrate the statement-level semantics contained in the global hierarchy, we show two classical examples:
\ding{182} The semantics of statements is related to the global hierarchy.
Most programming languages permit the creation of blocks and nested blocks. The block structure is fundamental for creating control flow and defining the scopes of variables. Thus modeling block structure is helpful for understanding control flow and variable scope. The control flow will influence the effects of a statement.
Figure \ref{fig:block} shows an example about semantics of the statement in the block structure. The statement \definenode{print(sum)} can be placed at any of the three marked positions: in the inner \definenode{for} loop, in the outer \definenode{for} loop, and outside of the outer \definenode{for} loop. A small change in the statement's position will affect the program's output.
Locating the statement in the block structure will help the model better determine the function of the source code.
\ding{183} We also observe that the functionality of the program is closely related to the global hierarchy.
Source code with similar functionalities tends to have similar global hierarchies. 
This unique global hierarchy can help the model distinguish the functionality and semantics of the program. 
To better confirm our observation, we conducted a simple statistical experiment on the code classification task on \textbf{Python800} dataset from the CodeNet project \cite{puri2021codenet}.
Figure \ref{fig:class} gives an example solution with problem id \textit{p02412} in CodeNet.
This program counts the number of triplets of numbers satisfying two requirements: each number is less than $n$ and they sum up to $k$.
There is a \definenode{if} statement in four levels of nesting while/for loop. We traverse the dataset and find that there are 121 programs that have a \definenode{while-for-for-for-if} hierarchical position. We surprisingly find that 111 of the 121 (about 91.7\%) programs are written to solve problem \textit{p02412}. And there are 300 programs in total for solving this problem, which means about 37\% of the programs solving \textit{p02412} use the special structure mentioned above.
Through these statistics, we claim that the global hierarchy of source code is strongly related to the functionality and semantics of the program.

Through our analysis, we show that global and local hierarchical structures are essential for code representation models, while existing joint learning models ignore the former. We will conduct an empirical study to prove our point experimentally in Section \ref{sec:globalandlocal}.



\begin{figure}[t]
\centering
  \includegraphics[width=\columnwidth]{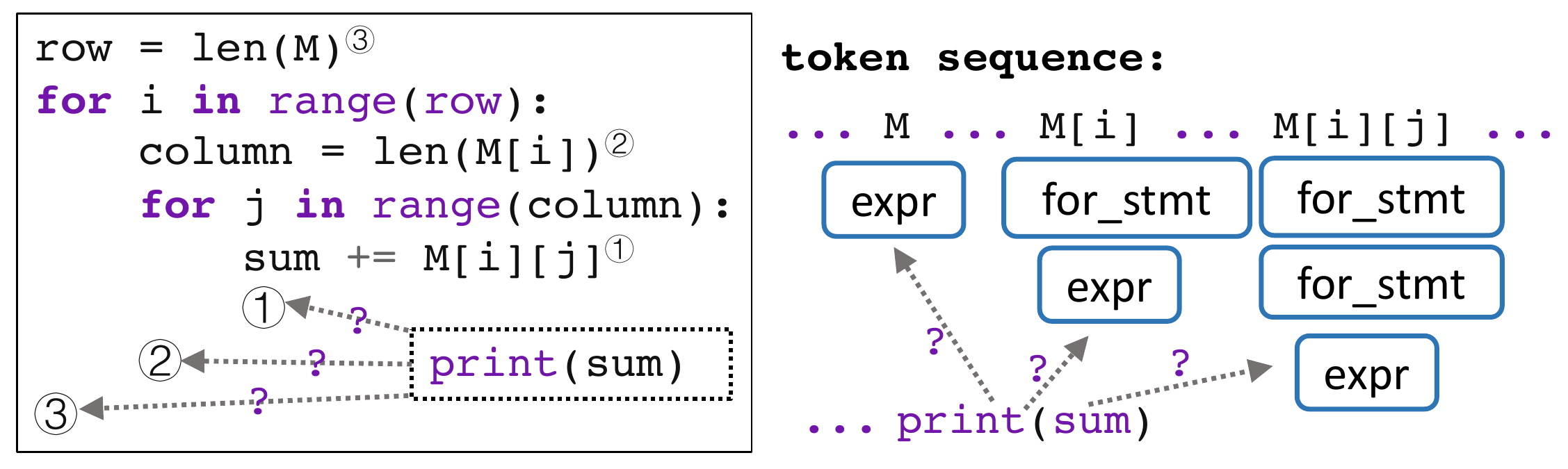}
\caption{An illustrative example of the implicit block structure ignored in token sequences. We can place the statement \definenode{print(sum)} in three places, where the token sequences are almost the same, but the semantics are different.}
\label{fig:block}
\vspace{-15pt}
\end{figure}

\begin{figure}[t]
\centering
  \includegraphics[width=\columnwidth]{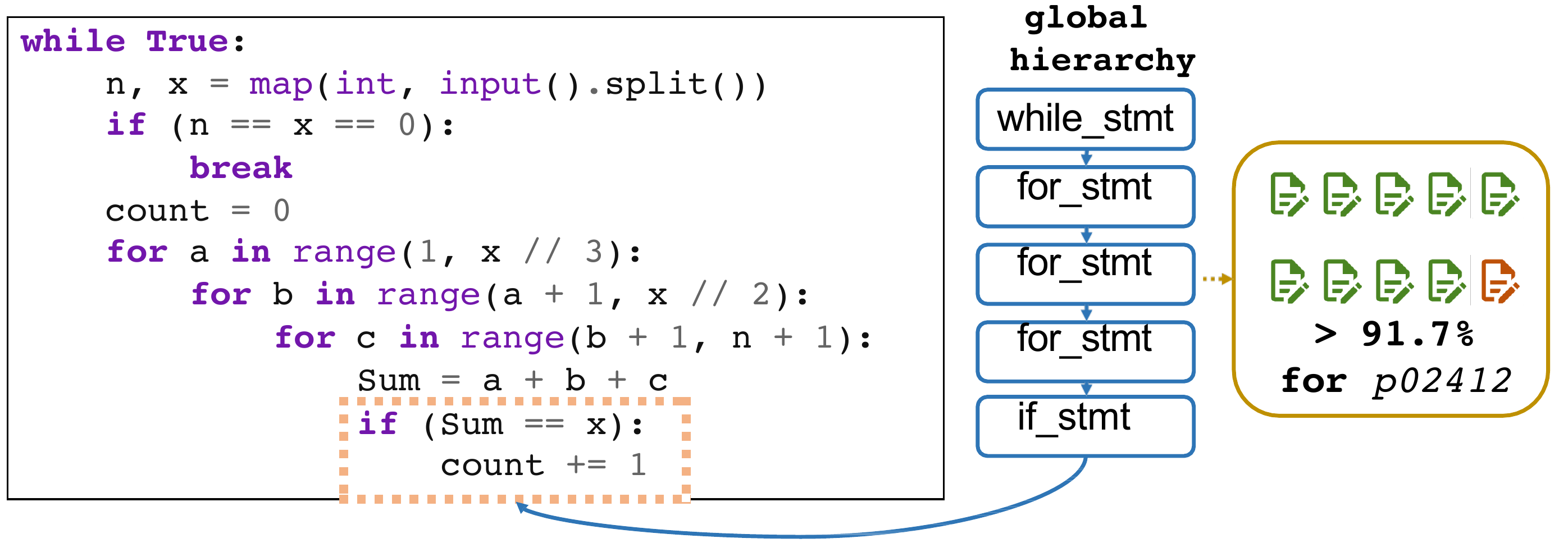}
\caption{An illustrative solution example from problem id \textit{p02412} in CodeNet Python800. There are 121 programs that have such a hierarchical position in the dataset. 111 of the 121 programs (more than 91.7\%) are written for the problem \textit{p02412}.}
\label{fig:class}
\vspace{-15pt}
\end{figure}









\section{Proposed Model}
\label{sec:model}
\begin{figure*}[t]
\centering
  \includegraphics[width=2\columnwidth]{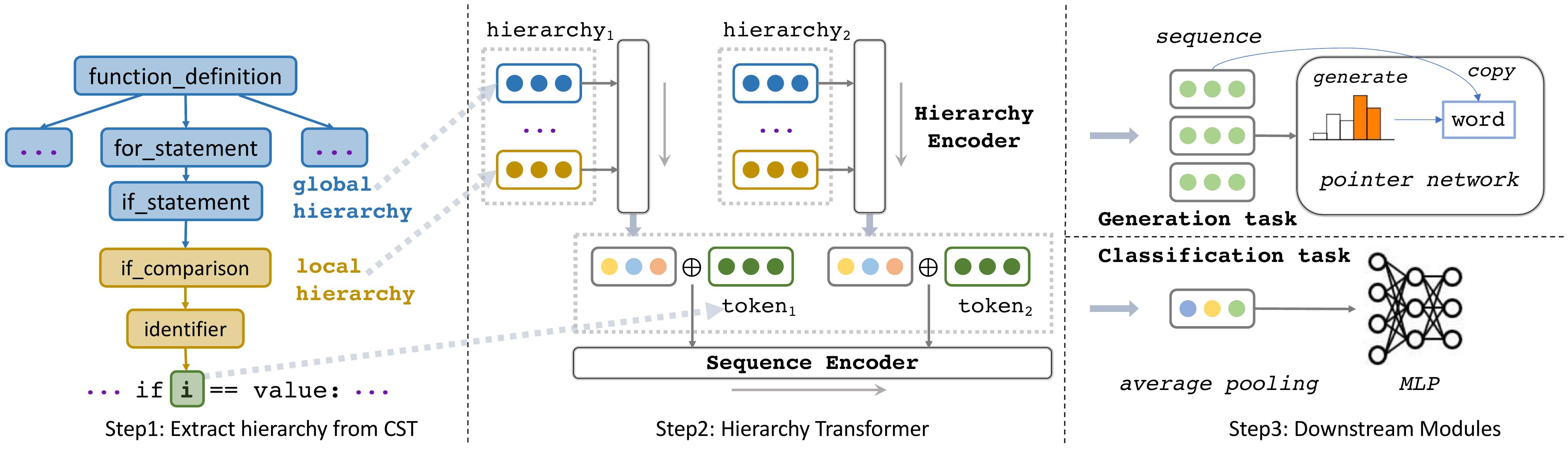}  
\caption{The pipeline of our approach}
\label{fig:pipeline}
\vspace{-15pt}
\end{figure*}

\subsection{Overview}
In this work, we propose a source code representation model, HiT, to encode the code sequence and the hierarchy information simultaneously. The entire pipeline of our approach is shown in Figure \ref{fig:pipeline}. To get the hierarchical position of the source code, we first parse the source code to the concrete syntax tree and extract paths from it.
Each path is fed into a Transformer-based hierarchy encoder to embed the hierarchy information to hidden vectors. 
We then concatenate the hierarchy vector representation with the token embedding and use another Transformer-based sequence encoder to learn the final code representation.
Finally, the representation vector can be fed into a linear classifier or a decoder for various downstream tasks.

\subsection{Hierarchy Extraction}
%
Hierarchical embeddings in  source code are extracted from the concrete syntax tree. 
Parsing trees of source code contains rich structural information of source code. 
Concrete syntax tree (CST) and abstract syntax tree (AST) are parsing trees. The concrete syntax tree reflects the exact syntax of the source code, where each leaf node corresponds to a source code token. In theory, a CST can be converted to an AST equivalently. However, ASTs do not represent every detail appearing in the real syntax. For instance, the braces, semicolons, and parentheses are discarded in ASTs, making it hard to align the structural information with these code tokens. To extract the hierarchical embedding of each source code token, we use CSTs in this work.

As stated in Section \ref{sec:motivation}, we expect to model the hierarchical embedding of source code to better understand the semantics of tokens and statements. To get the hierarchical embedding, we extract all root-to-leaf paths from the CST. Considering that each leaf node in the CST corresponds to a source code token, the extracted root-to-leaf paths can be aligned to each token. The set of paths expresses the hierarchy of the program.
Intuitively, we can further divide a path into two parts: the root-to-statement path and the statement-to-leaf path.
The root-to-statement path represents the surrounding block structure of a statement. We name the structure \textit{global hierarchy} which reflects the position of a statement in the program. The statement-to-leaf path represents the structure of the local context of a source code token. We name the structure \textit{local hierarchy} which reflects the position of a token within the statement.
\textit{Step 1} in Figure \ref{fig:pipeline} gives an illustration of different parts of the tree path.
The \textit{global hierarchy} and \textit{local hierarchy} are included in the root-to-leaf path. In our main experiments, we use the root-to-leaf path in HiT model. To study the contribution of \textit{global hierarchy} and \textit{local hierarchy}, respectively, we perform an empirical study where only one part of the path is used in RQ1 in Section \ref{exp:rq2}.

\subsection{Hierarchy Transformer}
We propose a new sequence model, Hierarchy Transformer (HiT), which uses a combination of token sequences and hierarchy information for code representation. 
Given a token sequence with hierarchical path $s = [(\textbf{n}_1, t_1) ,\cdots, (\textbf{n}_l, t_l)]$, where $l$ is the length of the token sequence, $(t_1,\cdots,t_l)$ indicate token sequence of the source code, and $\textbf{n} = (n_1,\cdots,n_{l'})$ is a tree path (sequence of tree nodes).
To process both token sequences and hierarchy information, we use a small Transformer to encode the paths $\textbf{n}_k$. We then concatenate the representation of hierarchy information and token embedding and feed them into a larger Transformer model. We refer to the two transformers as Hierarchy Encoder and Sequence Encoder.

\paragraph{(i) Hierarchy Encoder}
The hierarchy encoder is designed to process the hierarchy into a vector representation. For each tree path, we first embed the types of tree nodes and feed them into the Transformer model. Then we perform a mean pooling to get the representation of the whole path.

\begin{align}
    e_1, e_2, \cdots, e_{l'} = Embed(n_1, n_2, \cdots, n_{l'}) \\
    h_1, h_2, \cdots, h_{l'} = Transformer_{hie}(e_1, e_2, \cdots, e_{l'}) \\
    p = MeanPooling(h_1, h_2, \cdots, h_{l'})
\end{align}

\paragraph{(ii) Sequence Encoder}
The sequence encoder is designed to combine the hierarchy representation with tokens and process the resulting new sequence into the final code representation vector.
We first concatenate the hierarchy representations $p_i$ with the token embeddings $e_i$.
Then we use another Transformer as the sequence encoder to get the final code representation.

\begin{align}
    E_1, E_2, \cdots, E_l = Embed(t_1, t_2, \cdots, t_l) \\
    \mathbf{X} = (p_1 || E_1, p_2 || E_2, \cdots, p_l || E_l) \\
    H_1, H_2, \cdots, H_l = Transformer_{seq}(\mathbf{X})
\end{align}

\subsection{Downstream Modules}

The hierarchy encoder transforms the code sequence with hierarchy information into a vectorized representation. Most program processing tasks can be categorized into classification tasks and generation tasks. To apply our HiT on downstream tasks, we use different downstream modules according to the type of tasks.  

\paragraph{(i) Classification}
For classification tasks, models are required to classify programs based on the functionalities or other properties they implement. 
We first apply average pooling over the HiT output and get the global representation vector $v$.
After getting $v$, we apply a 2-layer MLP as the classifier to get the classification result. The probability of the output label is then calculated with a softmax layer:
\begin{gather*}
    v = MeanPooling(H_1, H_2, \cdots, H_l) \\
    o = g(W_2 \cdot f(W_1 \cdot v + b_1) + b_2) \\
    P_i = \frac{\mathop{exp}(o_i)}{\sum_i \mathop{exp}(o_i)}
\end{gather*}
We use the standard cross entropy loss to train our model:
\begin{equation}
    \mathcal{L} = -\sum_{i=1}^{|\mathcal{Y}|}\mathbbm{1}_{y==i}\log P_{i},
\end{equation}
where $\mathbbm{1}$ is the indicator function.

\paragraph{(ii) Generation}
For generation tasks, models are required to generate the target sequence conditioned on the encoder output, such as method name prediction and code summarization.
We pass all encoder output as a sequence to a transformer decoder.
To generate out-of-vocabulary (OOV) tokens, we adopt a pointer network \cite{vinyals2015pointer} based on the Transformer decoder model.
The pointer model first attends to the encoder's output at each timestep and gets a hidden vector $h^*_t$. 

\begin{gather*} 
    e^{t}_i = W_3^{T}\mathrm{tanh}(W_1 H_i + W_2s_t + b) \\
    a^t = \mathrm{softmax}(e^t) \\
    h^*_t =\sum_ia_i^{t} H_i,
\end{gather*}
where $W_1, W_2, W_3, b$ are learnable parameters, $s_t$ represents the output of the transformer decoder at timestep $t$.
After obtaining the context vector, the model produces the vocabulary distribution and the copy probability $p_{copy} \in [0,1]$ with $H_i$ and $s_t$ at this timestep. The $p_c$ denotes the probability of copying tokens from the input sequence. On the contrary, $p_{gen} = 1 - p_{copy}$ indicates the probability of generate a token from the vocabulary. The probability of predicting the token $w$ is calculated as follows:
\begin{gather*}
    P_v = \mathrm{softmax}(W_4h^*_t+b_1) \\
    P_{copy} = \mathrm{sigmoid}(W_5h^*_t+b_2) \\
    P(w) = p_{gen}P_{v}(w) + p_{copy}\sum_{i:w_i=w}a_i^{t}.
\end{gather*}
The copying mechanism enables the model to enhance its predictions by pointing at positions in the input sequence.
During training, the loss for the output sequence is calculated as the average loss over the negative log likelihood of each target token $w_t$:
\begin{align}
    \mathcal{L} = \frac{1}{T}\sum_{t=0}^T -\log \sum_{\tilde{w}_t\in vocab} \mathbbm{1}_{\tilde{w}_t==w_t}P(\tilde{w}_t)
\end{align}

\section{Experimental Setup}
\label{sec:experiment}
With the extracted hierarchy information in the code sequence, we adopt HiT and perform extensive evaluation upon three code understanding tasks involving classification and generation tasks across 8 different datasets. We aim to investigate five research questions:

\paragraph{RQ1. Global Hierarchy \& Local Hierarchy}
What is the impact of the global and local hierarchy on code representation models?
To what extent do the different types of hierarchy improve the performance of HiT?

\paragraph{RQ2. Scope Information In Global Hierarchy}
Can HiT learn the scope information in global hierarchy?
Is it important for the code representation model to focus on the global hierarchy?

\paragraph{RQ3. HiT vs. Transformer on Performance and Efficiency} 
Is the hierarchy information in HiT helpful for sequence models? What is the parameter cost and training efficiency of HiT compared with the vanilla Transformer?

\paragraph{RQ4. Performance on Classification Tasks}
How does HiT perform compared with the SOTA models on the code classification and clone detection tasks?

\paragraph{RQ5. Performance on Generation Tasks}
How does HiT perform on generation tasks? Can HiT produce better results than SOTA models on the method name prediction task?

\subsection{Subject tasks and Datasets}
Our experiments are conducted upon two representative source code classification tasks (\ie, code classification and clone detection) and one generation task (\ie, method name prediction). We evaluate 8 widely used datasets in total, and the compared baseline models are among the classical models or the SOTA models. The statistics of these datasets are summarized in Tabel \ref{tab:dataset_statistic}.

\begin{table*}[t]
\setlength\tabcolsep{5pt}
  \centering
  \caption{Statistics of datasets}

\begin{tabular}{clccccccccccc}
\toprule
\multicolumn{1}{l}{}      &                 & \multicolumn{5}{c}{Code Classification}                                                                                                &                      & \multicolumn{2}{c}{Clone Detection}     &                      & \multicolumn{2}{c}{Method Name Prediction}                         \\ \cline{3-7} \cline{9-10} \cline{12-13} 
\multicolumn{1}{l}{}      &                 & \multirow{2}{*}{Java250} & \multirow{2}{*}{Python800} & \multirow{2}{*}{C++1000} & \multirow{2}{*}{C++1400} & \multirow{2}{*}{POJ-104} &                      & \multicolumn{2}{c}{POJ-Clone}           &                      & \multirow{2}{*}{CSN-Ruby} & \multirow{2}{*}{CSN-Python} \\ \cline{9-10}
\multicolumn{1}{l}{}      &                 &                          &                            &                          &                          &                          &                      & Examples & \multicolumn{1}{l}{Problems} & \multicolumn{1}{l}{} &                           &                             \\ \midrule
\multirow{3}{*}{\rotatebox{90}{Size}}     & Train           & 45,000                   & 144,000                    & 300,000                  & 252,000                  & 36,400                   & \multicolumn{1}{l}{} & 32,000   & 64                           & \multicolumn{1}{l}{} & 48,791                    & 412,178                     \\
                          & Valid           & 15,000                   & 48,000                     & 100,000                  & 84,000                   & 5,200                    & \multicolumn{1}{l}{} & 8,000    & 16                           & \multicolumn{1}{l}{} & 2,209                     & 23,107                      \\
                          & Test            & 15,000                   & 48,000                     & 100,000                  & 84,000                   & 10,400                   & \multicolumn{1}{l}{} & 12,000   & 24                           & \multicolumn{1}{l}{} & 2,279                     & 22,176                      \\ \midrule
\multirow{5}{*}{\rotatebox{90}{Avg. Length}} & Token Seq       & 228.02                   & 125.18                     & 270.96                 & 334.89                   & 246.96                   &                      & \multicolumn{2}{c}{246.96}              &                      & 79.18                  & 131.59                    \\
                          & Complete Hierarchy & 9.26                     & 7.66                       & 7.14                     & 7.62                     & 9.15                     &                      & \multicolumn{2}{c}{9.15}                &                      & 6.89                & 9.12                  \\
                          & Global Hierarchy   & 7.15                     & 4.05                       & 4.75                     & 5.06                     & 6.22                     &                      & \multicolumn{2}{c}{6.22}                &                      & 5.62                  & 5.69                    \\
                          & Local Hierarchy   & 2.11                     & 3.61                       & 2.39                     & 2.56                     & 2.92                     &                      & \multicolumn{2}{c}{2.92}                &                      & 1.26                  & 3.42                    \\
                          & Target Seq      & -                        & -                          & -                        & -                        & -                        & \multicolumn{1}{l}{} & \multicolumn{2}{c}{-}                   &                      & 2.23                  & 2.25                    \\ \bottomrule
\end{tabular}
  \label{tab:dataset_statistic}
\vspace{-15pt}
\end{table*}

\subsubsection{\bf Code Classification}
In the code classification task, the model needs to predict the category of the given code snippet based on the semantics. In particular, for selected datasets, solutions under each question correspond to a category. We consider using Project CodeNet \cite{puri2021codenet} and POJ-104 \cite{tbcnn}.
Project CodeNet contains over 14M code samples from two open judge platforms AIZU and AtCoder. It provides four large and challenging datasets for the code classification task, including \textbf{Java250}, \textbf{Python800}, \textbf{C++1000} and \textbf{C++1400}.
\textbf{POJ-104} is collected from another pedagogical online judge system with 104 programming problems. It has been used by many previous studies in code classification and clone detection tasks.

\subsubsection{\bf Clone Detection}
In the clone detection task, the model needs to detect whether two pieces of code implement the same functionality.
We adopt the \textbf{POJ-Clone} and follow the previous task settings \cite{LuGRHSBCDJTLZSZ21codexglue}.
It aims to retrieve other programs that solve the same problem given a program. To test the generalization ability of different approaches, the training/validation/test is split based on the problems.

\subsubsection{\bf Method Name Prediction}
In method name prediction, a method with its name masked is fed into the model, and the model needs to predict the original method name based on the given method body. We experiment on two datasets introduced in the CodeSearchNet (CSN) Challenge \cite{husain2019codesearchnet,ZugnerKCLG21codetrans}: \textbf{CSN-Python} and \textbf{CSN-Ruby}. The datasets are obtained by scraping from public repositories across the most popular projects on GitHub.
We follow the work of \citea{husain2019codesearchnet} which splits the data based on the source repositories.


\subsection{Evaluation Metrics}

\paragraphwithoutdot{For code classification task}, we adopt the measure in \cite{puri2021codenet} and use the accuracy for this multiclass classification task. 

\paragraphwithoutdot{For clone detection task}, we follow \citea{LuGRHSBCDJTLZSZ21codexglue} and use the $MAP@R$ \cite{MusgraveBL20MAP} score for evalutation. 
$MAP@R$ is defined as the mean of the average precision scores, each of which is evaluated to retrieve the most similar $R$ samples given a query. 
For a code (query), R is the number of other codes in the same class and $R=499$ in the POJ-Clone dataset.

\paragraphwithoutdot{For method name prediction task}, we adopted metrics in previous studies\cite{code2vec,alon2018code2seq,ZugnerKCLG21codetrans}, which measure \textit{precision}, \textit{recall}, and \textit{f1} on subtokens of generated method names.

\subsection{Implementation Details}
\paragraph{Hierarchy Extraction Parser}
We extract the hierarchy information from the concrete syntax tree with \textit{Tree-sitter}, a parser generator tool. 
Based on this tool, our approach is general and dependency-free enough to parse any programming language and extract the hierarchy information. In our experiments, we have selected datasets in \textit{C++, Java, Python, Ruby} for evaluation, showing the generality and effectiveness.
\paragraph{Model Implementation}
Our model is implemented based on the Pytorch framework. We conduct all experiments on a Tesla V100S GPU with 32GB of memory. Each experiment is run five times with random seeds and then averaged for final results. We set the embedding size and the hidden size to 256, and employ 8 heads in each transformer layer. For the code classification task and the clone detection task, the hierarchy encoder consists of 2 layers, and the sequence encoder consists of 6 layers. For the method name prediction task, the number of sequence encoder layers and decoder layers are set to 4 and 2. We use AdamW with a learning rate of $1e^{-4}$ and weight decay. We use spaces as the separator for tokenizer, and set the vocabulary size between 5000-8000 according to different tasks.
For all baseline models, we retrain on the given datasets to get more reliable results. We try to keep the hyper-parameters the same as baseline models for a fair comparison.

\section{Experimental Results}
\label{sec:results}

\subsection{RQ1: Global vs. Local Hierarchy}
\label{sec:globalandlocal}
\label{exp:rq2}

\begin{table}[t]
    
    \centering
    \caption{Performance of HiT vs. Transformer, and HiT with different types of hierarchy for all three tasks.}
\centering
\subtable[{\normalsize On Code Classification and Clone Detection Tasks}]{
\centering
\setlength\tabcolsep{2pt}
\begin{tabular}{llccccclc}
\toprule
             &       & \multicolumn{5}{c}{Code Classification {\footnotesize (Accuracy)}}                                   & \multicolumn{1}{c}{\multirow{2}{*}{\begin{tabular}[c]{@{}c@{}}POJ-Clone\\ {\footnotesize (MAP@R)}\end{tabular}}}
            \\ \cline{3-7} 
             & Para  & {\footnotesize Java250}        & {\footnotesize Py800}      & {\footnotesize C++1000}        & {\footnotesize C++1400}        & {\footnotesize POJ}      & \multicolumn{1}{c}{}                         \\ \midrule
\textbf{HiT} & 4.55M & \textbf{94.81} & \textbf{95.97} & \textbf{95.05} & \textbf{93.27} & \textbf{97.08}  & \multicolumn{1}{c}{\textbf{80.46}} \\
Trans  & 4.50M & 93.49          & 93.99          & 89.93          & 67.87          & 88.13           & \multicolumn{1}{c}{67.15}          \\ \midrule
\textit{global}  & 4.55M & 93.79          & 94.84          & 91.35          & 83.90          & 94.56           & \multicolumn{1}{c}{74.85}          \\
\textit{local}  & 4.55M & 93.95          & 95.42          & 92.64          &90.45          & 96.37           & \multicolumn{1}{c}{75.88}          \\ \bottomrule
\end{tabular}
}
\subtable[{\normalsize On Method Name Prediciton Task}]{
\centering
\setlength\tabcolsep{4.5pt}
\begin{tabular}{llccclccc}
\toprule
            &        & \multicolumn{3}{c}{CSN-Ruby} &  & \multicolumn{3}{c}{CSN-Python} \\ \cline{3-5} \cline{7-9} 
            & Para   & P        & R       & F1      &  & P        & R        & F1       \\ \midrule
\textbf{HiT}         & 36.75M & \textbf{30.70}     & \textbf{27.58}   & \textbf{29.06}   &  & \textbf{37.25}    & \textbf{33.75}    & \textbf{35.41}    \\
Trans & 34.89M & 24.26    & 19.66   & 21.71   &  & 32.71    & 27.63    & 29.96    \\ \midrule
\textit{global}         & 36.75M & 24.90     & 22.89   & 23.87   &  & 34.26    & 29.29    & 31.58    \\
\textit{local}         & 36.75M & 28.69     & 25.58   & 27.05   &  & 35.34    & 30.50    & 32.74    \\ \bottomrule
\end{tabular}
}
   \label{tab:hierarchy}

\vspace{-20pt}
\end{table}

To investigate the impact of different types of hierarchy information, 
we conduct an empirical study and feed HiT with the global hierarchy and local hierarchy, respectively. We perform the analysis on all datasets of three tasks. The experimental results are listed in Table \ref{tab:hierarchy}. The row \textit{global} and \textit{local} refers to the two types of hierarchy.

On the basic program understanding task of the code classification task, 
we observe that the local hierarchy outperforms the global hierarchy by 0.16\%, 0.58\%, 1.29\%, 6.55\%, and 1.81\% in the five datasets. 
It shows that the local hierarchy is comparable and sometimes more effective than the global hierarchy for code classification. 
The \textit{MAP@R} of the local hierarchy on the clone detection task is improved by 1.03 compared with the global hierarchy.
For the method name prediction task, the local hierarchy outperforms the global hierarchy by 3.18 and 1.16 in F1-score on two datasets.

Experiments show that both the global and local hierarchy outperform the original sequence model, indicating that they are helpful for the sequence model to distinguish the semantics and functionalities of the programs.
The global hierarchy provides the surrounding block structure of a statement, which contains the nested structure and the global position of a statement in the code. And the local hierarchy provides the structure of the local context of a source code, which contains the type information and the local position of a token in the statement. 
Experimental results show that the type information and the local structure play significant roles in the code representation, which also confirms that the existing models can achieve good results based on token-level local hierarchy.

Furthermore, We also notice that the performance of using global hierarchy and local hierarchy separately is worse than using the complete hierarchy. 
\textbf{Therefore, we can show that both global and local hierarchical embeddings are essential for code representation.} We use the complete hierarchy to feed HiT in our follow-up experiments.


\greysum{\textbf{Answer to RQ1:} The empirical study upon classification and generation tasks demonstrates the impact of different types of hierarchical embeddings. The local hierarchy focuses on the type information and local position of a token and plays a significant role in the code representation.
Using both types of hierarchical embeddings together in HiT can further enhance the model performance.}

\subsection{RQ2: Scope Information in Global Hierarchy}
\label{sec:scope_detection}
To investigate the scope information in global hierarchy learned by HiT, we design a new task called \textbf{Variable Scope Detection}. Given two variables in a program, the model needs to detect whether these two variables are in a scope.
\footnote{The scope of a variable is a block structure in the entire program where the variable is declared, used, and can be modified.} 
Formally, given the representation vector of the two variables $h_A, h_B$, the probability $p_{{scope}_{A,B}}$ of variable $A$ and $B$ in the same scope is calculated by dot product following a sigmoid function:
\begin{equation}
    p_{{scope}_{(A,B)}} = \mathrm{sigmoid}(h_AW_sh_B)
\end{equation}
where $W_s$ is a learnable parameter.
We provide examples of the task in Table \ref{tab:probe_statistics}.
Considering that the variable scope is a mapping of the hierarchical structure information on variable tokens, this task requires the model to learn the accurate global hierarchical block structure for sequence tokens. 

\begin{table}
\begin{minipage}{0.5\linewidth}
\raggedright
    \centering
    \caption{Examples for Variable Scope Detection in C++ dataset}
\centering

\begin{tabular}{l|l}
\toprule
Example Program                                                                                                           & Variable Pair   \\ \midrule
\multirow{4}{*}{\begin{tabular}[c]{@{}l@{}}\textit{if (i \% 2 == 0)} \\   \quad  \textit{C -= B;} \\ \textit{else} \\  \quad   \textit{A -= D;}\end{tabular}} & (\textit{C}, \textit{B})             \\
                                                                                                                  & \textit{Same scope}      \\ \cline{2-2} 
                                                                                                                  & (\textit{C}, \textit{A})             \\
                                                                                                                  & \textit{Different scope} \\ \bottomrule
\end{tabular}

\label{tab:probe_statistics}
\end{minipage}\hfill
\begin{minipage}{0.45\linewidth}
\raggedleft
    \centering
    \caption{Performance of models on variable scope detection  (Accuracy)}
\centering

\begin{tabular}{lll}
\toprule
Model & Python          & C++             \\
\midrule
HiT         & \textbf{80.27} & \textbf{76.12} \\
\midrule
Transformer      & 77.18          & 63.77          \\
GGNN            & 79.81          & 68.94         \\
GREAT       & 79.39          & 69.38 \\
\bottomrule
\end{tabular}

\label{tab:probe}
\end{minipage}

\vspace{-15pt}
\end{table}

In our experiment, we adopt Python800 and C++1400 datasets in Project CodeNet. We sample variable pairs and extract about \textit{7 million} pairs for Python and \textit{65 million} pairs for C++ with balanced labels. The division of the dataset follows the settings suggested by CodeNet in the classification task. We extract the variable representation vector from the encoder output of trained models in classification task. For sequence models, we extract the corresponding token representation. For graph-based models, we extract from node representations. We compare with the vanilla Transformer, GGNN and GREAT. The selection details of baselines are discussed in Section \ref{sec:compare_seq_structure}. Experiment results are shown in Table \ref{tab:probe}.

Results show that with the hierarchical embedding, our model can understand the hierarchical block structure better and outperforms the vanilla Transformer and GGNN on the variable scope detection task. We can also observe that HiT performs even better on the C++ dataset by at least 6.7\%. Our in-depth investigation on the dataset reveals that the programs in the C++ dataset are much longer and the relationship between variables is more complex compared with the Python dataset. The program graphs in the C++ dataset for tree-based and graph-based models are large. The number of nodes in the receptive field of each node grows exponentially, which leads to these models not being able to understand the complete sequence information well. \textbf{Our model retains the advantage of the sequence models to capture long-term semantic dependency, and shows the importance of learning the scope information in global hierarchy.}

\greysum{\textbf{Answer to RQ2:} We design the variable scope detection task to explore the scope information learned by code representation models.
Experiments show HiT can retain the advantage of sequence models to capture long-term semantic dependency and enhance sequence models with the scope information in global hierarchy.}

\subsection{RQ3: HiT vs. Transformer on Performance and Efficiency}


To answer RQ3, we compare the hierarchy transformer with the vanilla Transformer. 
We list the results of our HiT and the vanilla transformer on all three tasks in Table \ref{tab:hierarchy}. We also show the parameter cost under different task settings. The column $Para$ refers to the total number of parameters for different models. In the following experiments we will continue to use the same amount of parameters.

\begin{figure}[t]
 \subfigure[Python800] {
  \label{fig:python800Training}     
  \includegraphics[width=0.45\linewidth]{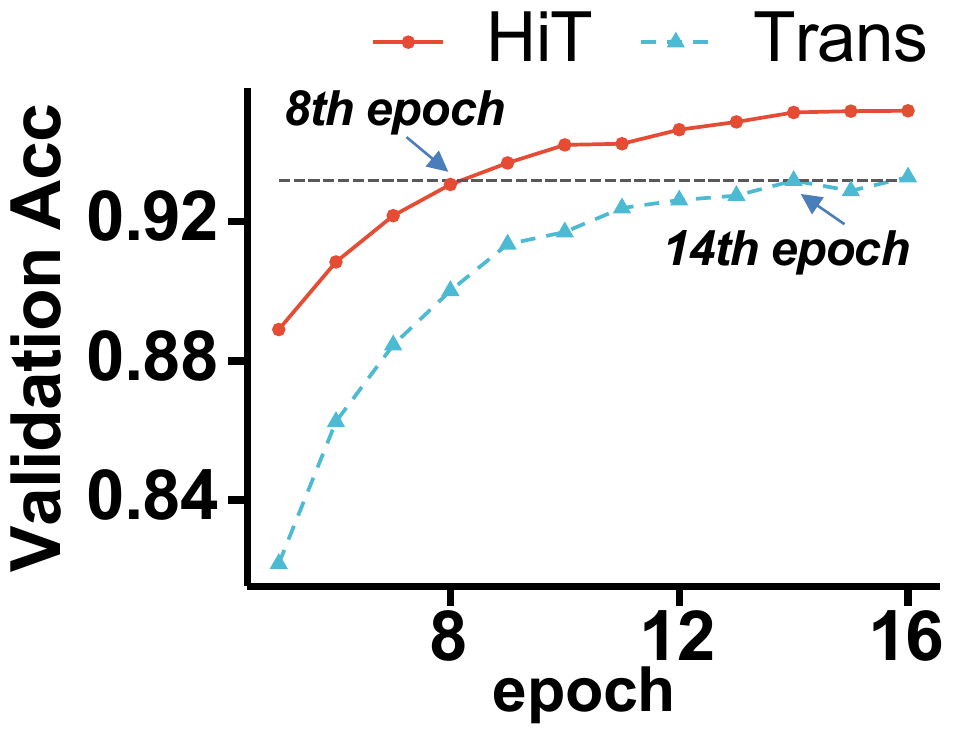}  
  }%
 \subfigure[C++1400] {
  \label{fig:c1400Training}     
  \includegraphics[width=0.45\linewidth]{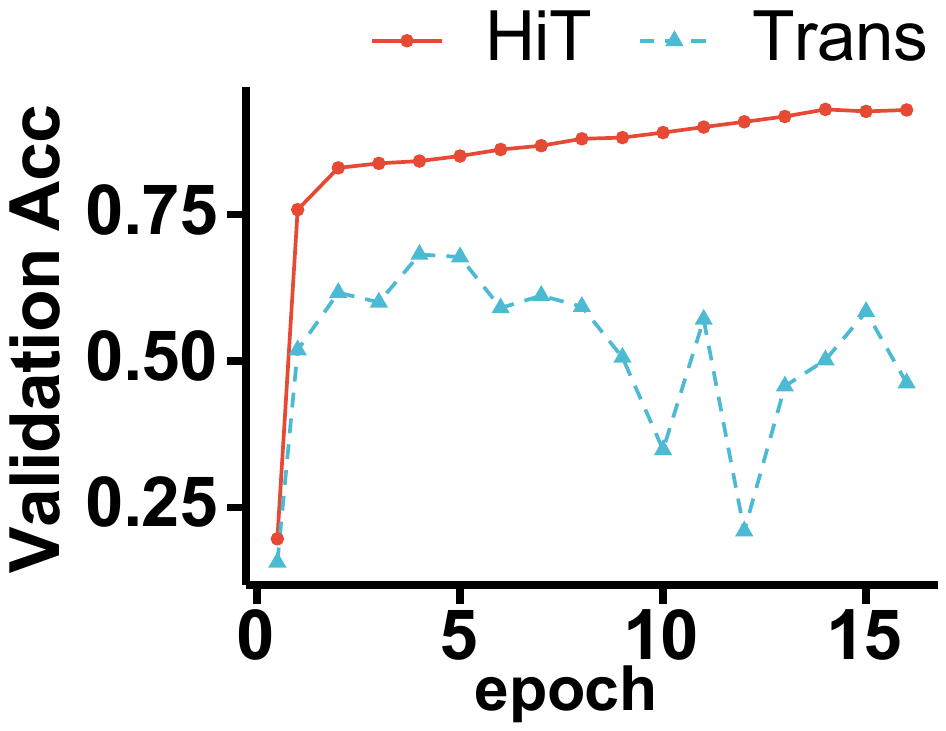}  
  }
\caption{Training efficiency of \textit{HiT} and \textit{Transformer} on Python800 and C++1400 datasets. In \ref{fig:python800Training} We plot a line  to show that \textit{HiT} can achieve comparable results to \textit{Transformer} with less training cost. In \ref{fig:c1400Training} it indicates that our model can achieve more stable training process on difficult datasets.}
\label{fig:codenetTraining}
\vspace{-10pt}
\end{figure}

For the classification tasks, we find that on C++1400 dataset, adding the hierarchy information would cause the classification accuracy to improve by more than 25.4\%. On C++1000 dataset, this improvement is more than 5.12\%. It shows that the hierarchy information can significantly improve the stability and performance of sequence models, especially for those datasets with complex semantics and long programs.
For the clone detection task, we observe that HiT outperforms the Transformer by 13\% on \textit{MAP@R}. Considering our experimental setup that training and testing sets are split according to OJ problems, HiT is more generalizable on the unseen problems than the original sequence model. For the method name prediction task, upon the two datasets CSN-Ruby and CSN-Python, the F1-score improves by about 7.35\% and 5.45\%.
This improvement is significant in both classification and generation tasks over 8 different datasets. Hence, the hierarchy information is essential for sequence models to generate accurate code representations for various tasks. Our model significantly enhances the original sequence model on different tasks, at a minimal extra parameter cost of 1\%-5\%.

To evaluate the training efficiency, We recorded the training process of the two models in Figure \ref{fig:codenetTraining}. We selected two representative datasets: Python800 and C++1400. 
The training process of our model is faster and more stable. On Python800 dataset, HiT can achieve comparable results to Transformer with less training epochs and time. For difficult datasets such as C++1400, HiT shows more stable learning ability, making it perform more efficiently on such datasets.

\greysum{\textbf{Answer to RQ3:} For both classification and generation tasks, the hierarchy transformer broadly gains better performance with little parameter cost compared with the vanilla Transformer, especially 
on datasets with complex semantics such as C++1400. HiT also achieves stable training efficiency compared with sequence models.}

\subsection{RQ4: On Classification Task}

\subsubsection{\bf RQ4.1: Code Classification}
\begin{table}[t]
    \setlength\tabcolsep{5pt}
    \centering
    \caption{Performance of models on code classification task}
\centering
\begin{threeparttable}
\begin{tabular}{lccccc}
\toprule
Model       & Java250 & Python800 & C++1000        & C++1400        & POJ-104 \\ \midrule
RGCN        & 91.93   & 91.60      & 92.73          & 92.34          & 95.57   \\
GGNN        & 93.64   & 92.23     & 91.72          & 92.48          & 94.80    \\ \midrule
TBCNN       & 92.84   & 93.17     & 94.77          & 88.29          & 96.20    \\
ASTNN       & 92.86   & 93.80     & 94.61          & 90.17          & 96.79     \\
TreeCaps    & 93.07   & 94.35     & 94.92          & 90.16          & 96.81   \\ \midrule
SBT         & 65.64   & 71.69     & 65.05          & 56.33          & 89.58   \\
X-SBT       & 83.31   & 89.10      & 66.79          & 67.00          & 94.58   \\ \midrule
Transformer & 93.49   & 93.99     & 89.93          & 67.87          & 88.13  \\
GREAT       & 93.36   &  93.27      & 92.76          & 92.50          & 90.33  \\ \midrule
\textbf{HiT}         & \textbf{94.81}   & \textbf{95.97}     & \textbf{95.05} & \textbf{93.27} & \textbf{97.08}   \\ \midrule
\textit{CodeBERT}\tnote{$\dagger$}       & \textbf{\textit{96.47}}   &  \textbf{\textit{97.41}}      & \textit{86.13}           &  \textit{83.05}          & \textbf{\textit{98.40}}  \\ 
\bottomrule
\end{tabular}
\begin{tablenotes}
\footnotesize
     \item[$\dagger$] CodeBERT is a pre-trained model with significantly more parameters than our model, with a parameter count nearly 27 times that of HiT.
\end{tablenotes}
   \label{tab:res_cls}
\end{threeparttable}
\vspace{-15pt}
\end{table}
We compare with the following state-of-the-art code representation models:
\baseline{(1) Graph Neural Networks} We include RGCN (\cite{SchlichtkrullKB18rgcn}) and GGNN (\cite{li2015ggnn}) as baseline models. 
\baseline{(2) Tree-structured Neural Networks} We include TBCNN \cite{tbcnn}, ASTNN \cite{icseZhangWZ0WL19astnn} and TreeCaps \cite{aaaiBuiYJ21treecaps} as baseline models. 
\baseline{(3) Traversal Sequences of ASTs} We compare our model with SBT \cite{hu2018deep}  and XSBT \cite{SPTCode}. 
SBT\cite{hu2018deep} represents the tree nodes into a token sequence with brackets to denote hierarchies. X-SBT \cite{SPTCode} simplifies SBT by using an XML-like form. 

\baseline{(4) Transformer-based Models} In addition to the vanilla transformer, we also compare our model with GREAT\cite{hellendoorn2019global}. 
For comparison, we also include results for CodeBERT, a widely used pre-trained model that has a much larger number of parameters than HiT. Specifically, CodeBERT has 125 million parameters, nearly 27 times the number in our model.

The results in Table \ref{tab:res_cls} show that our model achieves the best performance on the code classification task over all baseline models across different program languages. The overall average accuracy on four datasets in Project CodeNet is at least 1.65\% higher than other baseline models, and the accuracy on POJ-104 dataset is increased by at least 0.27\% compared with the SOTA models. 
We observe that the graph-based or tree-based models are more effective than the vanilla Transformer. It proves that using only the token sequence, the sequence model cannot learn code representation well. Our proposed HiT addresses this issue for sequential models. With the enhancement of hierarchy information, HiT outperforms those graph/tree-based models. 
We also notice that directly feeding the traversal sequences of ASTs in the Transformer performs poorly. The flattened AST node sequence impairs the sequential information of the context in the source code. In comparison, our approach is much more efficient and effective in combining naturalness and hierarchy for code representation.
Even compared to large-scale models such as CodeBERT, our model is still competitive. HiT achieves comparable performance and even outperforms on C++1000 and C++1400 datasets by 8.92\%, 10.22\%. The overall average accuracy for Project CodeNet of HiT is 4.01\% higher than CodeBERT. We must be aware that CodeBERT requires a large model size and pre-training on over 8 million data samples (while our model can be trained on a single GPU and does not require pre-training). 
We also notice sequence models often struggle on C++1000 and C++1400 datasets, including pre-trained models. Further investigations are conducted in Section \ref{sec:analysis_C++}.


\subsubsection{\bf RQ4.2: Clone Detection}
\begin{figure}[t]
\centering
  \includegraphics[width=0.8\linewidth]{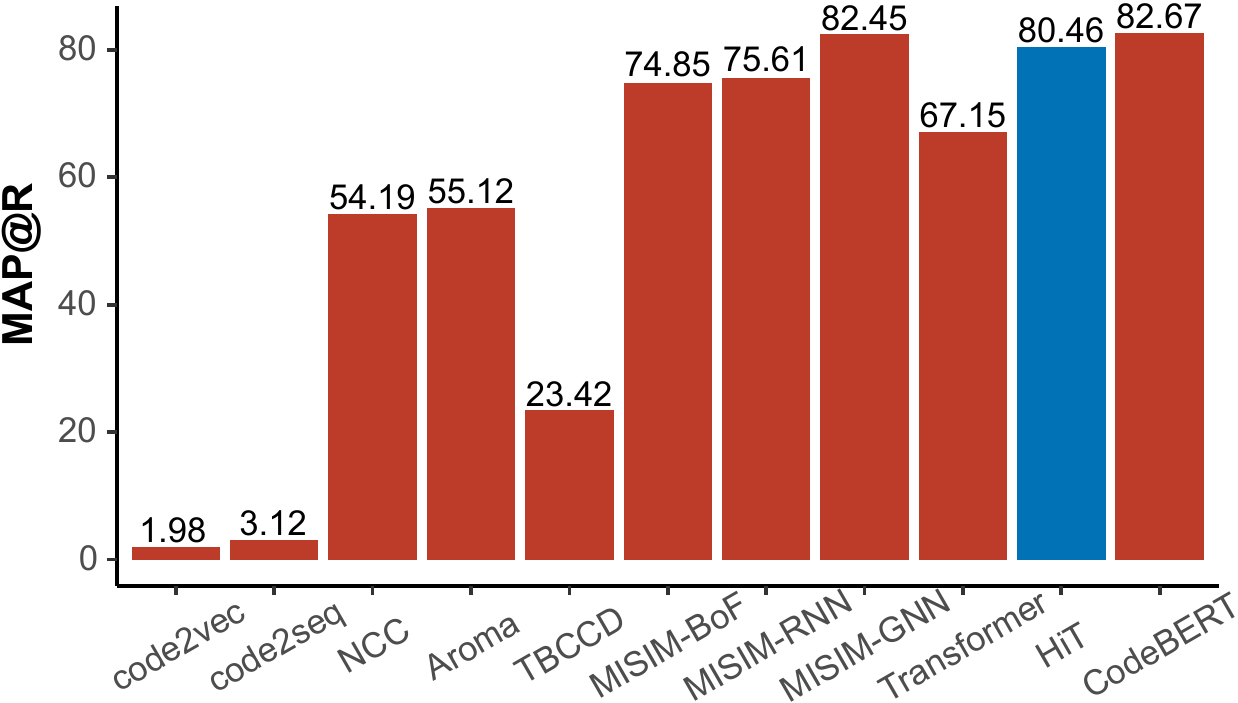}  
\caption[Caption for LOF]{Performance of models on POJ-Clone.\footnotemark}
\label{fig:clone_rst}
\vspace{-15pt}
\end{figure}
\footnotetext{The results of baselines are from the CodeXGLUE benchmark.}
To further verify the generalization ability of the model in distinguishing program semantics, we choose to evaluate our model on a clone detection dataset (POJ-Clone) for further evaluation.
As we mentioned, the clone detection dataset is partitioned into training, validation, and test sets with different OJ problems.
We compare our model with several state-of-the-art methods specially designed for code clone detection related tasks.

\baselinewithoutdot{(1) Code2vec/Code2seq \cite{code2vec,alon2018code2seq}} uses the attention-based method with leaf-to-leaf paths of AST to learn embeddings of codes. 

\baselinewithoutdot{(2) NCC \cite{NunJH18NCC}} encodes programs by leveraging both the underlying data flow and control flow of the programs with LSTM to build a code similarity system.

\baselinewithoutdot{(3) Aroma \cite{LuanYBS019aroma}} is a code recommendation engine with the simplified parse tree (SPT).


\baselinewithoutdot{(4) TBCCD \cite{YuLCLXW19tbccd}} is a clone detection model with tree-based convolution networks.
It achieves state-of-the-art performance on a simplified version dataset of clone detection. \footnote{In the authors' paper, they evaluate their model on a clone detection dataset that is not partitioned based on OJ problems. The program semantics in the training and test sets are the same. However, we think this setting weakens the generalization ability of the model.}

\baselinewithoutdot{(5) MISIM \cite{MISIM} } is a code clone detection system that incorporates the context-aware semantics structure (CASS) in its design. This structure has been carefully tailored to support the analysis of specific programming languages.

We list the results in Figure \ref{fig:clone_rst}. In most cases, our HiT outperforms among baseline models, which improves by at least 25.34 in \textit{MAP@R} except MISIM models. 
The MISIM models need to be evaluated on every possible combination of manually designed configurations \cite{MISIM}, thus, the preprocessing process is complex. Our method is simple and easy to use, with comparable results with the best MISIM-GNN among those models. When conducted on sequence models, our performance is even better than MISIM-RNN. 

We observe that in our challenging experimental setting, TBCCD performs poorly. We also notice that NCC, Aroma and MISIM both require complex preprocessing designed for particular languages. Our model is based on CSTs and is more generalizable and practical.
Even compared with the extremely large pretrained models, our model achieves the comparable performance on the clone detection task. 


\greysum{\textbf{Answer to RQ4:} Our proposed model HiT generally outperforms the current SOTA baseline models upon code classification and clone detection tasks. 
Our proposed HiT can understand program semantics and handle the classification tasks well. }

\subsection{RQ5: On Generation Task}
\begin{table}[t]
    \centering
    \caption{Results of models on method name prediction task}
\centering
\begin{tabular}{lccccccc}
\toprule
\multicolumn{1}{c}{\multirow{2}{*}{Model}} & \multicolumn{3}{c}{CSN-Ruby} &  & \multicolumn{3}{c}{CSN-Python} \\ \cline{2-4} \cline{6-8} 
\multicolumn{1}{c}{}                       & P        & R       & F1      &  & P        & R        & F1       \\ \cline{1-4} \cline{6-8} 
Code2seq                                   & 23.23    & 10.31   & 14.28   &  & 35.79    & 24.85    & 29.34    \\
GGNN                                       & 19.15    & 14.11   & 16.24   &  & 24.07    & 19.09    & 21.29    \\
SBT                                        & 19.84    & 12.22   & 15.12   &  & 30.92    & 18.32    & 23.01    \\
X-SBT                                      & 22.82    & 13.04   & 16.60    &  & 34.58    & 20.69    & 25.89    \\ \midrule
Transformer                                & 24.26    & 19.66   & 21.71   &  & 32.71    & 27.63    & 29.96    \\
GREAT                                      & 24.66    & 22.25   & 23.39   &  & 35.09    & 31.62    & 33.26    \\ \midrule
CodeTrans                            & \textbf{31.46}    & 24.50    & 27.55   &  & 36.41    & 33.68    & 34.99    \\
GTNM                                       & 24.59    & 20.11   & 22.13   &  & 32.98    & 27.73    & 30.13    \\ \midrule
\textbf{HiT}                                       & 30.70     & \textbf{27.58}   & \textbf{29.06}   &  & \textbf{37.25}    & \textbf{33.75}    & \textbf{35.41}    \\ \bottomrule
\end{tabular}

   \label{tab:res_gen}
\vspace{-15pt}
\end{table}
To answer this RQ, in addition to the baselines mentioned, we also compare with the SOTA models on the method name prediction task, including CodeTransformer \cite{ZugnerKCLG21codetrans} and GTNM \cite{liu22GTNM}. 
\baselineinline{
(1) CodeTransformer} learns structure and context jointly, and achieves state-of-the-art performance on CSN datasets.
\baselineinline{(2) GTNM} is a latest transformer-based model for method name prediction which extracts local contexts and project-level contexts and incorporates them into the sequence model.
The original paper of GTNM uses contexts designed for Java that are not available for CSN-Ruby and CSN-Python. We use a variant of GTNM that only considers the local context for fairly comparison.
In this experimental setup, we did not cover the tree models included in Table \ref{tab:res_cls}, as we found that they did not perform well on this task. We use the same copy mechanism of HiT for all baseline models.

We list the results of our HiT with the baselines upon CSN-Ruby and CSN-Python in Table \ref{tab:res_gen}.
The experimental results show that our HiT outperforms the baseline models significantly upon CSN-Ruby and CSN-Python. Specifically, our model outperforms CodeTransformer by 1.5071, 0.4216 in F1-score, respectively. And our model also significantly outperforms GTNM by 6.93, 5.28. 

\greysum{\textbf{Answer to RQ5:} Our proposed approach HiT outperforms all classic code representation models and current SOTA baselines on CSN-Ruby and CSN-Python method name prediction tasks. It suggests that HiT may be capable of producing better code representations for code-to-text generation tasks in general programming languages.}

\section{Discussions}
\label{sec:discuss}
\subsection{Comparison with existing sequence/structure-based models}
\label{sec:compare_seq_structure}

Due to the design of the model structure, the existing sequence model (\eg, the vanilla Transformer \cite{vaswani2017attention}) or structural-based model (\eg, ASTNN \cite{icseZhangWZ0WL19astnn}, TBCNN \cite{tbcnn}, GGNN \cite{li2015ggnn}) is usually better at capturing the information of a certain modal. Some studies jointly learn both sequential and structural information for code representation in sequence-based models such as GREAT \cite{hellendoorn2019global}. They focus on modeling structure as a relation between tokens with attention mechanism and overlook the full impact of hierarchical structure, especially for the global hierarchy. In this paper, we comprehensively investigate the impact of different types of hierarchy information on code representation tasks. In Section \ref{sec:scope_detection}, we design a Variable Scope Detection task to check the ability of different models to capture the global hierarchy information. We choose one of the sequence models, structural-based models and jointly learning models as the baseline. Experiments show that our model can better identify the information brought in the hierarchy. Our model retains the advantage of the sequence models to capture long-term semantic dependency, and shows the importance of implanting the full hierarchy information.

\subsection{Analysis of C++ datasets in Project CodeNet}
\label{sec:analysis_C++}
We further investgate the reason why the sequence model does not 9perform well on the C++ dataset.
Analysis of the code sequences in these C++ datasets revealed that they are longer (as shown in Table \ref{tab:dataset_statistic}). Upon further investigation, we treat the programs from the same class as a single text snippet by concatenating them and calculate the TF-IDF cosine similarity between every two classes \cite{Wang2022LearningPR}. The average similarity for C++1000 and C++1400 is 0.787 and 0.754, while the average score for Java250 and Python800 is only 0.722 and 0.420. This suggests that code sequences in C++ datasets are highly similar even in different classes, making it difficult for sequence models to accurately classify them. HiT shows strong training stability, especially on such difficult datasets.

\subsection{Time Efficiency of HiT}
In our experiments, when training HiT on the Python800 dataset, the additional pre-processing step only required $<5$ minutes to process 144000 samples, and HiT achieved the best performance after 122 minutes. In comparison, the vanilla transformer required 169 minutes. During inference, both models required approximately 30 seconds. This demonstrates the efficiency of HiT, especially at training time.

\subsection{Threats to Validity}
\label{sec:validity}
\paragraphwithoutdot{Threats to internal validity} relate to the roles of the model architecture and hyper-parameters setting.
In our experiments, we do a small-range grid search on learning rate and batch size settings. 
Another threat comes from the implementation of GTNM. We do not use additional project and document-level context for method name prediction as the original paper \cite{liu22GTNM} because they are not available in CSN-Ruby/Python.
\paragraphwithoutdot{Threats to external validity} mainly relate to tasks and datasets we choose. We counter this by evaluating our model on 8 different datasets of three tasks, including classification and generation tasks across 4 programming languages.

\paragraphwithoutdot{Threats to construct validity} include the evaluation metrics we used in this work. 
These metrics are adequate for corresponding tasks and have been adopted by many previous studies \cite{puri2021codenet,LuGRHSBCDJTLZSZ21codexglue,MusgraveBL20MAP,code2vec,alon2018code2seq,ZugnerKCLG21codetrans}.


\section{Conclusion}
\label{sec:conclusion}
In this paper, we analyze how the complete hierarchical structure influences tokens in code sequence representation and put forward the property of hierarchical embedding, including statement-level global hierarchy and token-level local hierarchy.
We propose HiT, a practical approach to incorporate hierarchical embedding into Transformer. 
We investigate the effectiveness of the global and local hierarchy with a detailed empirical study. The results show both hierarchies are essential for code representation models while existing joint learning models ignore the former.
Our in-depth evaluations demonstrate that HiT can generate accurate and delicate representations and outperforms the SOTA baselines for classification and generation tasks on 8 challenging datasets.

\section{Acknowledgement}

This research is supported by the National Natural Science Foundation of China under Grant No. 62072007, 62192733, 61832009, 62192730.
We also would like to thank all the anonymous reviewers for constructive comments and suggestions to this paper.


\bibliographystyle{ACM-Reference-Format}
\bibliography{sample-base}

\end{document}